\documentclass[12pt]{article}
\begin{document}

\begin{center}
{\large \bf Hypothesis about semi-weak interaction and experiments with solar neutrinos. II. Deuteron disintegration by neutral currents}

\vspace{0.3 cm}

\begin{small}
L.M.Slad\footnote{slad@theory.sinp.msu.ru} \\
{\it Skobeltsyn Institute of Nuclear Physics,
Lomonosov Moscow State University, Moscow 119991, Russia} \\
\end{small}
\end{center}

\vspace{0.3 cm}

\begin{footnotesize}
The present work provides one more evidence of that the solar neutrino problem has an elegant solution based on the hypothesis about the existence of a new, semi-weak, interaction. The analysis of the deuteron disintegration by neutral currents of solar neutrinos, generated by both the electroweak and semi-weak interactions, is fulfilled. A good agreement between the theoretical and experimental results for this process is obtained, which is in harmony with the conclusions of the first part of the work on the other four observed processes with solar neutrinos.
\end{footnotesize}

\vspace{0.3 cm}

\begin{small}

\begin{center}
{\bf 1. Introduction}
\end{center}

In the work \cite{1}, we have substantiated that the discrepancy between the predictions of the standard solar model for the rates of a number of processes caused by solar neutrinos and the results of the appropriate experiments testify for the existence of a new interaction, named by us semi-weak. We have demonstrated that the new interaction hypothesis, having only one free parameter, provides a good agreement between the theoretical and experimental characteristics of four out of the five observed processes with solar neutrinos: ${}^{37}{\rm Cl} \rightarrow {}^{37}{\rm Ar}$, ${}^{71}{\rm Ga} \rightarrow {}^{71}{\rm Ge}$, $\nu_{e} e^{-}\rightarrow \nu_{e} e^{-}$, and $\nu_{e}D \rightarrow  e^{-}pp$. The fifth observed process, which is the deuteron disintegration by neutral currents of solar neutrinos, $\nu_{e}+D \rightarrow \nu_{e}+n+p$, has remained not considered in \cite{1}. Its analysis, fulfilled in the paper work, again demonstrates a good agreement between the consequences of the new interaction theory and the experimental results. If in the results of the first four mentioned processes, the consequences of the semi-weak interaction of neutrinos with nucleons only inside the Sun are manifested, then in a result of the fifth named process, the consequences of such an interaction also in the terrestrial installation are manifested.

The semi-weak interaction is mediated by a massless pseudoscalar boson $\varphi_{ps}$. It is inherent at least to the electron neutrino, the proton and the neutron and is described by the following Yukawa Lagrangian
\begin{equation}
{\cal L} = ig_{\nu_{e}ps}\bar{\nu}_{e}\gamma^{5}\nu_{e}\varphi_{ps}+
ig_{Nps}\bar{p}\gamma^{5}p\varphi_{ps}-ig_{Nps}\bar{n}\gamma^{5}n\varphi_{ps},
\label{1}
\end{equation}
or by a similar Lagrangian with $u$- and $d$-quarks instead of proton $p$ and neutron $n$.

At every act of the elastic scattering of an electron neutrino with energy $\omega_{1}$ on a rest nucleon with mass $M$ caused by interaction (\ref{1}), the neutrino handedness changes from left to right or vice versa. In the first approximation, we assume that the fluxes of left- and right-handed electron neutrinos at the Earth's surface are approximately equal.

As a result of the elastic $\nu N$-scattering, the energy of the final neutrino $\omega_{2}$ becomes less than the initial one. It can take evenly distributed values in interval
\begin{equation}
\frac{\omega_{1}}{1+2\omega_{1}/M} \leq \omega_{2} \leq \omega_{1}.
\label{2}
\end{equation}

Since the energy of the solar neutrinos is less than about 18.8 MeV, then in the first approximation, we can with sufficient accuracy assume that the total cross section $\nu N$-scattering
\begin{equation}
\sigma = \frac{(g_{\nu_{e}ps}g_{Nps})^{2}}{16 \pi M^{2}}
\cdot \frac{1}{(1+2 \omega_{1}/M)},
\label{3}
\end{equation}
and, consequently, the number of collisions of neutrinos with the Sun nucleons does not depend on the initial energy $\omega_{1}$.

The only free parameter of the theory with a new interaction governing its consequences for the observed processes with solar neutrinos is the product of the Yukawa coupling constants from the Lagrangian (\ref{1}), $g_{\nu_{e}ps}g_{Nps}/4\pi$. In \cite{1}, we initially used the effective number $n_{0}$ of collisions of solar neutrinos with nucleons, occurring during the motion of neutrinos from the point of its production to the exit from the Sun, as a free parameter instead of the product of the Yukawa coupling constants. Its value $n_{0}=11$ corresponds to the best agreement. Then, based on this number, the following estimate was obtained
\begin{equation}
g_{\nu_{e}ps}g_{Nps} /4\pi = (3.2 \pm 0.2)\times 10^{-5}.
\label{4}
\end{equation}

The deuteron disintegration by neutral currents of solar neutrinos is caused by two non-interfering sub-processes, which differ in neutrino handedness either in the initial or in the final state. One sub-process, with the participation of only the left-handed neutrinos, is a standard one and is due to the exchange of the $Z$-boson. Another, unordinary sub-process with the participation both left- and right-handed neutrinos is due to the exchange of the massless pseudoscalar boson. As the postulated boson is isoscalar, its coupling constants with the proton and the neutron are opposite. Therefore, the cross section of the second sub-process is characterized by an additional factor $((M_{n}-M_{p})/M)^{2}$, so that it becomes comparable with the cross section of the first,  standard sub-process. It is quite natural that the effective solar neutrino flux corresponding to the total theoretical rate of the deuteron disintegration by the neutral currents, appears to be noticeably larger than the effective flux connected with the deuteron disintegration by the charged current.

At calculating the rate of the process of the deuteron disintegration by neutral current $\nu_{e}+D \rightarrow \nu_{e}+n+p$, and the corresponding effective flux of solar neutrinos, we use the procedures described in detail in the work \cite{1}. We present final results for the effective number of collisions $n_{0}=11$ and, for comparison, for the number $n_{0}=10$, as in \cite{1}.

\begin{center}
{\bf 2. Cross sections, rates and effective fluxes related to the process of the deuteron disintegration by neutral currents of solar neutrinos}
\end{center}

Experiments on the deuteron disintegration into a neutron and a proton caused by the neutral current of solar neutrinos are considered crucial for the theoretical interpretations. The consecutive results of such experiments in SNO are expressed with the following values of effective solar neutrino flux
$\Phi_{eff}^{nc}({}^{8}{\rm B})$ (in units of $10^{6}$ ${\rm cm}^{-2}{\rm s}^{-1}$): 
$$5.09^{+0.44}_{-0.43}{}^{+0.46}_{-0.43} \; \cite{2}, \quad 5.21 \pm 0.27 \pm 0.38 \; \cite{3}, \quad 4.94^{+0.21}_{-0.21}{}^{+0.38}_{-0.34} \; \cite{4}, \quad 5.54^{+0.33}_{-0.31}{}^{+0.36}_{-0.34} \; \cite{5}.$$ 
In view of our hypothesis about the existence of the interaction described by Lagrangian (\ref{1}), the deuteron disintegration into a neutron and a proton is described by two non-interfering sub-processes which differ in neutrino handedness either in the initial or in the final state. 

The first sub-process caused by left-handed neutrinos is standard, i.e. it is due to the exchange of the $Z$-boson. We use the tabulated values of the total cross section $\sigma^{{\rm nc}(Z)}(\omega)$ of this sub-process as function of the energy $\omega$ of the incident neutrinos, which are contained in the work \cite{6}. The rate of the sub-process of the deuteron disintegration due to $Z$-boson exchange $V(nc(Z)|{\rm B})$ is calculated according to a formula similar to (\ref{8}) of the work \cite{1}. This rate translates into effective flux of solar neutrinos with their spectrum from the decay of ${}^{8}{\rm B}$, $\Phi_{eff}^{nc(Z)}({}^{8}{\rm B})$, according to the formula
\begin{equation}
V(nc(Z)|{\rm B}) = \Phi_{eff}^{nc(Z)}({}^{8}{\rm B})
\sum_{i=1}^{160}\Delta^{B}p(\omega_{i}^{B})
\sigma^{{\rm nc}(Z)}(\omega_{i}^{B}), 
\label{5}
\end{equation}
where $\omega_{i}^{B}$ are the energy values of the neutrino from ${}^{8}{\rm B}$, which are given in the table of Ref. \cite{7} in the form of set $\omega_{i}^{B}=i\Delta^{B}$, where $i = 1, \ldots, 160$, $\Delta^{B}=0.1$ MeV, and their distribution is expressed through probability $p(\omega_{i}^{B})$ of that neutrinos possess energy in an interval $(\omega_{i}^{B}-\Delta^{B}/2, \ \omega_{i}^{B}+\Delta^{B}/2)$. We have
\begin{equation}
\Phi_{eff}^{nc(Z)}({}^{8}{\rm B}) = \left\{ 
\begin{array}{ll}
2.16 \cdot 10^{6} \; 
{\rm cm}^{-2}{\rm s}^{-1}, & {\mbox{\rm if}} \hspace{0.3cm} n_{0}=10, \\
2.10 \cdot 10^{6} \; 
{\rm cm}^{-2}{\rm s}^{-1}, & {\mbox{\rm if}} \hspace{0.3cm} n_{0}=11.
\end{array} \right.
\label{6}
\end{equation}

The second sub-process of the deuteron disintegration into a neutron and a proton, caused both left- and right-handed neutrinos, is due to the exchange of the massless pseudoscalar boson. When choosing the method and performing the calculation of the relevant cross section, we take into consideration a number of works on the deuteron disintegration by neutrinos, in particular, the works \cite{6}, \cite{8}--\cite{15}. We do not aspire to the degree of accuracy in the choice of the deuteron model and to the precision of calculation which have been achieved in Ref. \cite{6}, \cite{8}--\cite{10}. We are based on the spherically symmetric $np$-potential of the type $U_{0}\delta(r)$ used in Ref. \cite{11}--\cite{14}.

We denote the 4-momenta and their components in the laboratory frame as
\begin{equation}
\nu_{e}(k)+D(P_{0}) \rightarrow \nu_{e}(k')+N_{1}(p_{1})+N_{2}(p_{2}),
\label{7}
\end{equation}
\begin{equation}
k=\{ \omega, {\bf k}\}, \; k'=\{ \omega', {\bf k}'\}, \; P_{0}=\{M_{D}, 
{\bf 0}\}, \; p_{1}=\{ E_{1}, {\bf p}_{1}\}, \; p_{2}=\{ E_{2}, {\bf p}_{2}\}.
\label{8}
\end{equation}
We also introduce the 3-momentum of the center of mass ${\bf P}= {\bf p}_{1}+{\bf p}_{2}$ and the relative 3-momentum of the final nucleons ${\bf p}= ({\bf p}_{1}-{\bf p}_{2})/2$. We describe the deuteron and the final $N_{1}N_{2}$-system state vectors by tensor products (1) of the scalar functions of time and coordinates of the nucleons, and (2) of vectors from the spin space, where the latter are represented by the tensor product of two Dirac bispinors. In turn, the above scalar functions are given by the product (1) of the center-of-mass motion wave functions, and (2) of the functions of the distance $r$ between the nucleons, $\varphi_{D}(r)$ and $\varphi_{N_{1}N_{2}}(r)$. It is accepted that
\begin{equation}
\varphi_{D}(r) = \sqrt{\frac{\gamma}{2\pi}} \cdot \frac{e^{-\gamma r}}{r}, \quad
\varphi_{N_{1}N_{2}}(r) = \frac{\sin(pr+\delta_{0})}{pr}, 
\label{9}
\end{equation}
where $p=|{\bf p}|$, $\gamma = \sqrt{MB}$,  $p\cot\delta_{0} = -1/a_{s}$,  
$(a_{s}^{2}M)^{-1} = 0.0738$ MeV. It is easy to make sure of the following known equality
\begin{equation}
\left| \int d^{3}{\bf r}  \varphi_{D}^{*}(r)\varphi_{N_{1}N_{2}}(r) \right|^{2}
= \frac{8\pi \gamma (1-a_{s}\gamma)^{2}}{(1+a_{s}^{2}p^{2})
(p^{2}+\gamma^{2})^{2}}.
\label{10}
\end{equation}

Let us divide the final state phase space of the deuteron disintegration process into such pairs related to the protons and neutrons, so that the pseudoscalar current with interaction (\ref{1}) be represented in the form
$$J^{5} = \frac{g_{Nps}}{\sqrt{2}} \int d^{3}{\bf r}\left\{ \varphi_{D}^{*}(r)
\left[ \bar{\psi}({\bf p}_{a},M_{p})\gamma^{5}\psi({\bf p}'_{a},M_{p})\cdot
\bar{\psi}({\bf p}_{b},M_{n})\psi({\bf p}'_{b},M_{n}) \right. \right. $$
\begin{equation}
\left. \left. -\bar{\psi}({\bf p}_{a},M_{n})\gamma^{5}\psi({\bf p}'_{a},M_{n})\cdot \bar{\psi}({\bf p}_{b},M_{p})\psi({\bf p}'_{b},M_{p})\right] 
\varphi_{N_{1}N_{2}}(r)\right\}, 
\label{11}
\end{equation}
omitting the mass center motion.
Since the current (\ref{11}) as a function of the mass $M_{n}$ vanishes at $M_{n}=M_{p}$, it is advisable to expand the bilinear forms $\bar{\psi}({\bf p}_{c},M_{n}){\cal O}\psi({\bf p}'_{c},M_{n})$, $c=a,b$, in a Taylor series in powers of $M_{n}-M_{p}$. Restricting to the first power and to the approximation that $E_{i} = M$, we get
\begin{equation}
J^{5} = \frac{g_{Nps}}{\sqrt{2}} \cdot \frac{M_{n}-M_{p}}{M} 
\left[\bar{\psi}({\bf p}_{a},M_{p})\gamma^{5}\psi({\bf p}'_{a},M_{p})\cdot
\bar{\psi}({\bf p}_{b},M_{p})\psi({\bf p}'_{b},M_{p})\right]
\int d^{3}{\bf r} \varphi_{D}^{*}(r)\varphi_{N_{1}N_{2}}(r).
\label{12}
\end{equation}

The factor $1/\sqrt{2}$ in relations (\ref{11}) and (\ref{12}) is necessary to avoid double counting of the cross section due to taking each set of the final states in the phase space determined by the energy-momentum conservation. The phase space considering the kinematics of the deuteron disintegration (\ref{7}) as $2 \rightarrow 3$ process is given by the following formula
\begin{equation}
d^{3}I = \delta(\omega+M_{D}-\omega'-E_{1}-E_{2})\delta({\bf k}-{\bf k}'-
{\bf p}_{1}-{\bf p}_{2}) \; d^{3}{\bf k}' \; d^3{\bf p}_{1} \; d^3{\bf p}_{2}.
\label{13}
\end{equation}
We use the nonrelativistic approximation
\begin{equation}
E_{1}+E_{2}=M_{n}+M_{p}+\frac{{\bf p}_{1}^{2}}{2M}+\frac{{\bf p}_{2}^{2}}{2M}, 
\label{14}
\end{equation}
substitute ${\bf P}$ and ${\bf p}$ for ${\bf p}_{1}$ and ${\bf p}_{2}$ in equation (\ref{13}), and integrate over ${\bf P}$, eliminating the delta-function of 3-momenta. We obtain
\begin{equation}
d^{2}I =  \delta\left( \omega-B-\omega'-\frac{({\bf k}-{\bf k}')^{2}}
{4M}-\frac{p^{2}}{M}\right) \; d^{3}{\bf k}' \; d^3{\bf p}.
\label{15}
\end{equation} 

On the basis of relations (\ref{1}), (\ref{12}) and (\ref{10}), we find the squared modulus of the matrix element $|{\cal M}|^{2}$ for the sub-process of the deuteron disintegration, caused by the massless axial boson exchange between the left- and right-handed neutrinos and nucleons. Substituting it into the formula for the differential cross section
\begin{equation}
d\sigma = \frac{|{\cal M}|^{2}}{16\omega \omega' M^{2} (2\pi)^{5}}d^{2}I,
\label{16}
\end{equation} 
using equality (\ref{15}) with knowing the relative energy $E_{r}=p^{2}/M$ and performing a number of integrations, we obtain
$$\sigma^{\rm nc(ps)}(\omega) = \frac{(g_{\nu_{e}ps}g_{Nps})^{2}}
{16\pi^{2}M^{2}}\cdot \left( \frac{M_{n}-M_{p}}{M}\right)^{2}\cdot 
\frac{\sqrt{B} (\sqrt{B}-(a_{s}\sqrt{M})^{-1})^{2}}{\omega}$$
\begin{equation}
\times \int_{0}^{\omega - B} dE_{r}\frac{(\omega - B -E_{r})
\sqrt{E_{r}}}{(E_{r}+B)^{2}(E_{r}+(a_{s}^{2}M)^{-1})}.
\label{17}
\end{equation}

Using the value of the product of the coupling constants of the massless axial boson with the electron neutrino and nucleons given below by equality (\ref{4}) and the above values of other constants, we find the cross section values for the discussed deuteron disintegration sub-process for several energies of the incident neutrino $\omega = (2.2+0.2 j)$ MeV, $j=1,2,\ldots,70$. A number of these values is shown in table 5. To ensure that the results of calculations of the cross section $\sigma^{\rm nc(ps)}$ by formula (\ref{17}) are satisfactory, we have carried out calculations of the cross section $\sigma^{\rm nc(Z)}$ in an approach similar to that described above, and we have placed the results of them in the columns "Control" of table 5 along with the accurate enough results, taken from \cite{6}.

\newpage
\begin{center}
{\bf Table 1.} The cross section of the deuteron disintegration \\ sub-processes 
 $\sigma^{\rm nc(Z)}$ и $\sigma^{\rm nc(ps)}$ 
(in units of $10^{-42}$ cm$^{2}$), \\
\begin{tabular}{cccccccc}
\multicolumn{8}{c} {caused by $Z$-boson and massless axial boson.} \\ 
\hline
\multicolumn{1}{c}{$\omega$}
&\multicolumn{1}{c}{$\sigma^{\rm nc(Z)}$}  
&\multicolumn{1}{c}{$\sigma^{\rm nc(Z)}$}
&\multicolumn{1}{c}{$\sigma^{\rm nc(ps)}$} 
&\multicolumn{1}{c}{$\omega$}
&\multicolumn{1}{c}{$\sigma^{\rm nc(Z)}$}  
&\multicolumn{1}{c}{$\sigma^{\rm nc(Z)}$}
&\multicolumn{1}{c}{$\sigma^{\rm nc(ps)}$} \\
(MeV) & \cite{6} & Control & Eq. (\ref{17}) 
&(MeV) & \cite{6} & Control & Eq. (\ref{17}) \\
\hline
3.0  & 0.003 & 0.006 & 0.047 & 10.0 & 1.107 & 1.103 & 0.382 \\
4.0  & 0.031 & 0.039 & 0.132 & 11.0 & 1.458 & 1.443 & 0.402 \\
5.0  & 0.096 & 0.108 & 0.201 & 12.0 & 1.860 & 1.831 & 0.418 \\
6.0  & 0.203 & 0.217 & 0.255 & 13.0 & 2.314 & 2.268 & 0.433 \\
7.0  & 0.356 & 0.406 & 0.298 & 14.0 & 2.822 & 2.754 & 0.446 \\
8.0  & 0.557 & 0.568 & 0.331 & 15.0 & 3.382 & 3.290 & 0.457 \\
9.0  & 0.807 & 0.812 & 0.359 & 16.0 & 3.995 & 3.875 & 0.467 \\
\hline
\end{tabular}
\end{center}

We now are able to calculate the rate of the deuteron disintegration into a neutron and a proton, $V(nc(ps)|{\rm B})$, caused by the neutrinos due to the exchange of the massless axial boson with nucleons. Doing it by a formula similar to (\ref{8}) of the work \cite{1}, where we now put the total flux of left- and right-handed neutrinos $\Phi({}^{8}{\rm B})$ in place of the flux of left-handed neutrinos $0.5 \Phi({}^{8}{\rm B})$. After this, replacing the rate $V(nc(Z)|{\rm B})$ with the rate $V(nc(ps)|{\rm B})$ in the equality (\ref{5}) and keeping for the cross section $\sigma^{{\rm nc}(Z)}$ the status of theoretical cross section calculated on the basis of the standard model of electroweak interactions (as in Ref. \cite{6}), we find the effective solar neutrino flux $\Phi_{eff}^{nc(ps)}({}^{8}{\rm B})$ which is responsible for the rate of the deuteron disintegration  $V(nc(ps)|{\rm B})$. We have
\begin{equation}
\Phi_{eff}^{nc(ps)}({}^{8}{\rm B}) = \left\{
\begin{array}{ll}
(2.90 \pm 0.36) \cdot 10^{6} \; 
{\rm cm}^{-2}{\rm s}^{-1}, & {\mbox{\rm if}} \hspace{0.3cm} n_{0}=10, \\
(2.87 \pm 0.36) \cdot 10^{6} \; 
{\rm cm}^{-2}{\rm s}^{-1}, & {\mbox{\rm if}} \hspace{0.3cm} n_{0}=11,
\end{array}  \right.
\label{18}
\end{equation}
and the uncertainty in (\ref{18}) is only due to the uncertainty in the coupling constant (\ref{4}).

So, the effective flux of solar neutrinos corresponding to the total rate of the two sub-pro- cesses of the neutron disintegration by the neutral currents of the solar neutrinos, $V(nc(Z)|{\rm B})+V(nc(ps)|{\rm B})$, is equal to
\begin{equation}
\Phi_{eff}^{nc}({}^{8}{\rm B}) = \left\{ 
\begin{array}{ll}
(5.06 \pm 0.36) \cdot 10^{6} \; 
{\rm cm}^{-2}{\rm s}^{-1}, & {\mbox{\rm if}} \hspace{0.3cm} n_{0}=10, \\
(4.97 \pm 0.36) \cdot 10^{6} \; 
{\rm cm}^{-2}{\rm s}^{-1}, & {\mbox{\rm if}} \hspace{0.3cm} n_{0}=11,
\end{array} \right.
\label{32}
\end{equation}
that is in good agreement with the above-listed experimental results of SNO.

\begin{center}
{\bf 3. Conclusion}
\end{center}

It is justified to think that the range of phenomena potentially involved in the manifestation of the postulated semi-weak interaction will expand with time. In our opinion, it is first of all necessary to pay close attention to setting up an experiment on the deuteron disintegration by neutral currents of the reactor antineutrinos when placing the detector near a reactor. We expect that, due to the additional contribution of the semi-weak interaction, the observed rate of events in this experiment will be greater than the rate calculated on the basis of the standard model of electroweak interactions. 

I am sincerely grateful to S.P. Baranov, A.M. Snigirev, and I.P. Volobuev for useful discussions on some of the problems anyhow concerning the present work.

\end{small}
\end{document}